\documentclass[journal]{IEEEtran/IEEEtran}

\usepackage{cite}
\usepackage[pdftex]{graphicx}
\usepackage{amsmath}

\begin{document}

\title{Anomalous Subthreshold Behaviors in Negative Capacitance Transistors }

\author{Yu-Hung Liao,
        Daewoong Kwon,
        Suraj Cheema, 
        Ava J. Tan,
        Ming-Yen Kao,
        Li-Chen Wang,\\
        Chenming Hu,~\IEEEmembership{Life Fellow,~IEEE,}
        and Sayeef Salahuddin,~\IEEEmembership{Fellow,~IEEE,}
\thanks{Y.-H. Liao, D. Kwon, A. J. Tan, M.-Y. Kao, C. Hu and, S. Salahuddin are with the Department of Electrical Engineering and Computer Sciences, University of California, Berkeley, Berkeley, CA 94720, USA. (e-mail: yh\_liao@berkeley.edu).}
\thanks{S. Cheema and L.-C. Wang are with the Department of Material Science and Engineering, University of California, Berkeley, CA 94720, USA.}
\thanks{D. Kwon is now with the Department of Electrical Engineering, Inha University Yonghyeon Campus, Incheon 22212, South Korea}
\thanks{Y.-H. Liao and D. Kwon contributed equally to this work.}
\thanks{This work was supported by the Berkeley Center for Negative Capacitance Transistors.}}

\maketitle

\begin{abstract}

Recent measurements on ultra-thin body Negative Capacitance Field Effect Transistors have shown subthreshold behaviors that are not expected in a classical MOSFET.
Specifically, subthreshold swing was found to decrease with increased gate bias in the subthreshold region for devices measured over multiple gate lengths down to 30 nm.
In addition, improvement in the subthreshold swing relative to control devices showed a non-monotonic dependence on the gate length.
In this paper, using a Landau-Khanatnikov ferroelectric gate stack model calibrated with measured Capacitance-Voltage, we show that both these anomalous behaviors can be quantitatively reproduced with TCAD simulations.

\end{abstract}


\section{Introduction}

\IEEEPARstart{N}{egative} capacitance field effect transistors (NCFET) rely on a ferroelectric gate insulator to provide an amplification of the gate signal\cite{salahuddin2008use,salahuddin2008can,Krivokapic2017}. 
This boost, which depends on the capacitance matching between the ferroelectric and Si underneath, in turn, reduces the power supply voltage requirements. 
In a properly designed MOSFET, such a boost can also lead to a sub-thermal subthreshold swing (SS). 
However, the strong function of Si capacitance with voltage and inadequacy of available ferroelectric materials makes it difficult to obtain a sub-thermal subthreshold swing without hysteresis \cite{Wong2019,agarwal2019proposal}. 
On the other hand, the objective of reducing supply voltage can be achieved by improving the swing near the threshold and this can be done without violating the condition necessary for zero hysteresis \cite{khan2011ferroelectric}. 
NCFET so designed can show obvious non-classical behaviors.

\begin{figure}
    \centering
    \includegraphics{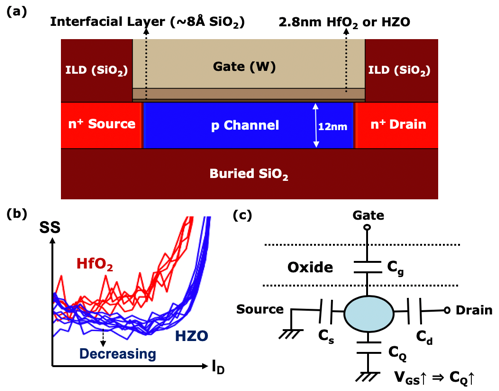}
    \caption{
    (a) Schematic cross-section of the SOI n-MOSFETs. Both HfO$_2$ (Control) and HZO (NC) devices have the same geometry.
    (b) SS of HfO$_2$ and HZO devices demonstrating conventional and anomalous trend with respect to $I_D$, respectively. (c) Equivalent circuit diagram for understanding the MOSFET channel barrier height in subthreshold regime.
    }
    \label{fig:schematics}
\end{figure}

Indeed, substantially improved I-V characteristics and non-classical subthreshold behavior have recently been observed in Zr doped HfO$_2$ gate stack (HZO) \cite{Kwon2020,Kwon2019,Chung2017,Yu2017,Zhou2016} that has a polar order \cite{Sang2015,Park2015}.
Despite the same thermal processing, doping, and MOSFET geometry (Fig.~\ref{fig:schematics}(a)), the HfO$_2$ (Control) and HZO (NCFET) gate stack devices demonstrate opposite trends of subthreshold swing with respect to gate bias (Fig.~\ref{fig:schematics}(b)) \cite{Kwon2020}. 
Figure \ref{fig:SS_Calib} shows that for all gate lengths ($L_G$) at $V_{DS}$=0.05V, the NCFET SS is similar to SS of the Control of the same geometry at low drain current ($I_D$=10pA/$\mu$m) but is lowered as the gate bias increases.
The $I_D$ at which NCFET subthreshold swing reaches its minimum is almost 4 orders of magnitude larger than the OFF current. 
In addition, the difference in the subthreshold swing between Control and NCFET devices (measured in the range $I_D$=0.1nA$\sim$1nA/$\mu$m) increases significantly from $L_G$=100nm to $L_G$=50nm but decreases from $L_G$=50nm to $L_G$=30nm (Fig.~\ref{fig:SStrend}).
Remarkably, these two trends for the HZO devices cannot be explained by assuming a ``higher-$\kappa$'' linear dielectric gate stack which would lead to monotonically increasing improvement of SS as $L_G$ shrinks.
Neither are they explainable with a better interface quality than the Control, as it would result in a constant SS reduction over different $L_G$.
In contrast with normal dielectric theory, these anomalies indicate that NCFET gate capacitance ($C_g$) (Fig.~\ref{fig:schematics}(c)) is affected significantly by both $V_{GS}$ and $L_G$, of which the explanation lies in the non-linear polarization response to the applied electric field for the HZO gate stack layer \cite{Hoffmann2019,Wong2019}.

\section{TCAD Model Calibrations}
Sentaurus TCAD \cite{Sdevice2018} simulator parameters are calibrated to C-V and I-V data from \cite{Kwon2020}. 
For both Control and NCFET, the gate insulator constitutes of a chemical oxide (8Å) and 2.8nm layer of HfO$_2$ or HZO.
HfO$_2$ MOSCAP Accumulation C-V measurements can be well matched by a 1.1 nm EOT gate stack in the TCAD model (Fig.~\ref{fig:CV_Vt}(a)). 
The HZO stack demonstrates a higher capacitance which could be fitted with very high dielectric constant for the HZO stack.
However, this would require an effective dielectric constant ($>$100) that is much higher than any theoretical predictions for Hf and Zr-based dielectric phases \cite{Rignanese2004,Zhao2005,Chen2011,Fischer2008,Gargi2006}.
In addition, this hypothetical ``super-high $\kappa$'' dielectric would not be able to replicate the anomalous subthreshold behaviors that we observed.
Instead, the capacitance boost is captured by the potential amplification effect \cite{Gao2014,Zhou2016,Hoffmann2019} induced by a the presence polar phase inside the HZO layer.

\begin{figure}
    \centering
    \includegraphics[width=3.5in]{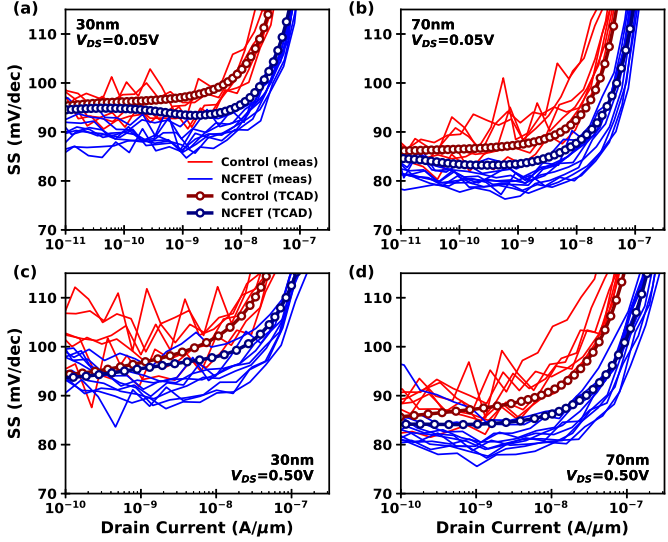}
    \caption{(a)-(d) Measured and TCAD-simulated SS-$I_D$ relations for Control and NC n-MOSFETs of two gate lengths. Two different drain biases are considered. Multiple devices are measured for each gate length, all of which exhibit negligible hysteresis.}
    \label{fig:SS_Calib}
\end{figure}

\begin{figure}
    \centering
    \includegraphics[width=3.5in]{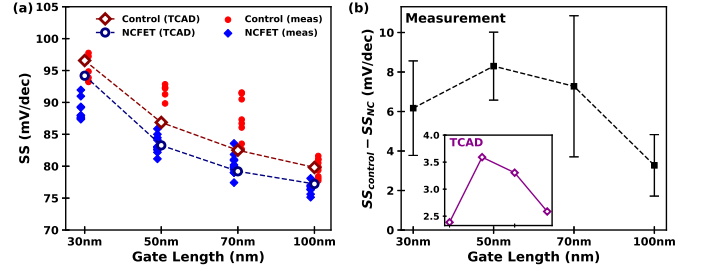}
    \caption{(a) SS averaged over $I_D$=0.1$\sim$1nA/$\mu$m at $V_{DS}$=0.05V for measured and TCAD-simulated Control and NC MOSFET. Each marker presents one device having the corresponding gate length. 
    (b) Experimentally estimated SS improvement for different gate lengths. Each error bar presents the estimated mean and one standard deviation. (Inset) TCAD-modeled Improvements for the four gate lengths. }
    \label{fig:SStrend}
\end{figure}

The calibrated simulation reproduces the experimentally observed subthreshold behavior for the Control MOSFET (Fig.~\ref{fig:SS_Calib},~\ref{fig:SStrend}(a)).
NCFETs are simulated by introducing the Landau-Khalatnikov (LK) model for simulating the gate stack, but non-gate-stack parameters are unchanged from the Control devices.
The LK parameters are chosen such that the Ferroelectric in the positive capacitance (PC) region results in the same gate stack EOT (1.1 nm) as the Control.
Therefore, the SS in the low $I_D$ (10-100pA/$\mu$m) region are matched between Control and NCFET devices.
As the gate bias is increased, the Ferroelectric enters into the negative capacitance (NC) region, which results in a reduced gate stack EOT of 0.9nm in accordance with the HZO C-V.
This transition from positive to negative capacitance regime successfully captures the near-threshold SS reduction (Fig.~\ref{fig:SS_Calib}) with respect to $I_D$ as well as the non-monotonic SS improvement trend with respect to $L_G$ (Fig.~\ref{fig:SStrend}(b) inset). 
This is discussed in more details later.

\section{Results and Discussions}

\begin{figure}
    \centering
    \includegraphics[width=3.5in]{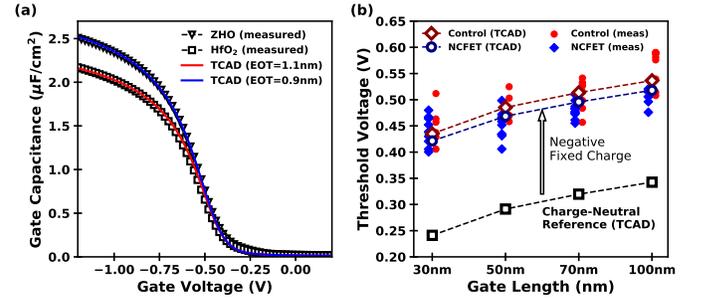}
    \caption{
    (a) Measured and TCAD-simulated C-V for p-type doped MOS capacitors with HfO$_2$ and ZHO gate stacks.
    (b) Constant-current ($I_D$=10nA/$\mu$m) threshold voltage (Vt) at $V_{DS}$=0.05V for measured and TCAD-simulated Control and NC MOSFET. Each marker represents an extraction for one device. Experimental $V_t$ are 0.18V larger than simulated $V_t$ without interface charge. The discrepancy is resolved when -0.6$\mu$C/cm$^2$ fixed charge is added to the Si/SiO$_2$ interface.}
    \label{fig:CV_Vt}
\end{figure}

Fig.~\ref{fig:CV_Vt}(b) shows threshold voltage ($V_t$) calibration results with tungsten work-function set to 4.6eV according to C-V calibrations (Fig.~\ref{fig:CV_Vt}(a)). Note that Control MOSFET $V_t$ would be underestimated by 0.18V if no net defect charges are assumed, and this discrepancy is insensitive to geometry and doping profiles when SS scaling trend is captured.
Therefore, the difference between the MOSCAP and MOSFET $V_{FB}$ is ascribed to additional defect charges introduced by transistor fabrication processes.
A fixed charge density of -0.6$\mu$C/cm$^2$ at the Si/SiO$_2$ interface can not only account for the $V_t$ shift but also explain the small measured $V_t$ difference between Control and NC devices (Fig.~\ref{fig:CV_Vt}(b)).
The voltage drop on the gate stack induced by the charges is less for the NCFET than the Control because of its larger $C_g$, which results in a lower $V_t$ that is consistent with the experiments. 
On the other hand, if there were no fixed charges, the NCFET $V_t$ would be larger than the Control $V_t$ because of mitigated short channel effects. 
Note that, the introduction of negative fixed charges into the model is consistent with our previous work as described in \cite{Liao2019} where the transistor fabrication followed essentially the exact same procedure. 

\begin{figure}
    \centering
    \includegraphics[width=3.5in]{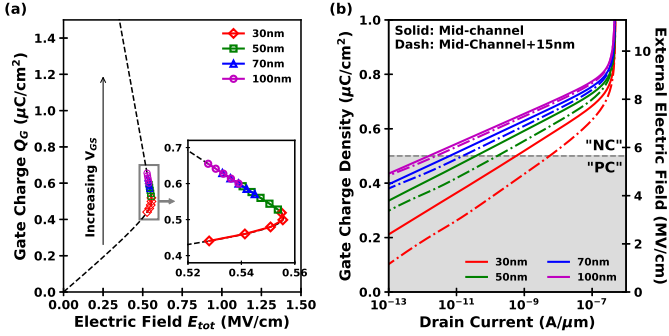}
    \caption{
    (a) Extracted Q-E relation for the polar layer of the gate stack at the mid-channel of NCFETs at $V_{DS}$=50mV. For each gate length, the gate voltage is ramped at which ID is from 0.1nA/$\mu$m to 1nA/$\mu$m.
    (b) Simulated relations between drain current and (left axis) local gate charge density / (right axis) local external electric field at $V_{DS}$=0.05V for NCFETs. For each gate length, the solid line is extracted at mid-channel gate stack region, and the dash line is extracted at the region 15nm laterally from mid-channel toward the drain. The white-background region corresponds to bias conditions at which the polar layer exhibits negative capacitance, while the dark background stands for positive-capacitance regime.
    }
    \label{fig:QI_QE}
\end{figure}

Fig.~\ref{fig:QI_QE} shows the transitions from positive to negative capacitance region of the HZO stack.
When $V_{GS}$ ramps up, both $I_D$ and the external electric field on the gate stack ($E_{ext}=Q_G/\epsilon_0$) increases.
The HZO polarization is directly determined by $E_{ext}$  if polarization gradient energy is negligible.
The $Q_G$-$I_D$ slope in the subthreshold regime is associated with input capacitance $C_{gg}=(\frac{1}{C_g}+\frac{1}{C_s+C_d+C_Q})^{-1}$, which increases for reduced $L_G$ because of $C_s$ and $C_d$ increase.
As the device enters the negative capacitance region, a voltage amplification ensues and drives down the subthreshold swing.
This explains why, for the NCFET, the subthreshold swing gets steeper in the sub-threshold regime as $V_{GS}$ or $I_D$ increases.
To fully understand the observed anomalous behaviors, especially the effect of $L_G$,one needs to consider the inner fringing field that plays a crucial role on the capacitance matching effect \cite{Liao2019,Lin2019}.
This field leads to the gate stack-to-source and gate stack-to-drain capacitive coupling which reduces $Q_G$ to balance the positively ionized source and drain donor charges for n-MOSFET in subthreshold regime.
When $L_G$ shrinks, mid-channel $Q_G$ corresponding to the same $I_D$ decreases because of the increased inner fringing field.
As a result, although the mid-channel gate stack is in the negative capacitance regime for $L_G$ from 50nm to 100nm at $I_D$=0.1nA/$\mu$m, it remains in the positive capacitance regime for the 30nm case (Fig.~\ref{fig:QI_QE}).
Therefore, the improvement in the subthreshold swing for NCFETs drops for $L_G$=30nm as shown in Fig.~\ref{fig:SStrend}(b).

Notably, the extracted $Q_G$ at which the HZO gate stack starts to exhibit negative capacitance is approximately 0.5$\mu$C/cm$^2$ (Fig.~\ref{fig:QI_QE}(a)).
In other words, the ‘S’ curve is pushed up in the charge axis.
This is a consequence of the negative fixed charge that simultaneously explains the $V_t$ trend for both Control and the NCFET.
The atypical response can be a result of antiferroelectric behavior of the tetragonal phase \cite{Park2015} in the HZO or a small but finite leakage that induces positive charge trapping at the SiO$_2$/HZO interface and effectively shifts the S-curve \cite{khan2017work,khan2016negative2}.

\section{Conclusion}

To summarize, we developed an FE model that simultaneously explains 
(i) SS getting steeper with increasing $I_D$ in the subthreshold regime, and 
(ii) non-monotonic behavior of SS improvement with $L_G$ – trends that cannot be explained by a classical ``high-$\kappa$'' scaling theory. 
Our results show that the capacitance matching leads to (i), while the inner fringe field induces change in the capacitance matching in short channel MOSFETs and leads to (ii).
This model supports the notion that Ferroelectric material optimization with appropriate device design \cite{Agarwal2018} to maximize near threshold capacitance matching may lead to significant reduction in $V_{DD}$. 

\clearpage
\bibliographystyle{IEEEtran}
\bibliography{IEEEabrv,Ref,Ref2}

\end{document}